\pdfoutput=1

 \documentclass[final,5p,times,twocolumn]{elsarticle}

 \usepackage{color}

 \usepackage{graphics}

 \usepackage{amsthm}

 \usepackage{lineno}

\biboptions{sort&compress,comma}

\usepackage{booktabs}

\usepackage[usenames, dvipsnames]{xcolor}
\usepackage[colorlinks = true]{hyperref}

\usepackage{etoolbox}

\makeatletter
\patchcmd{\ps@pprintTitle}{\footnotesize\itshape
       Preprint submitted to \ifx\@journal\@empty Elsevier
       \else\@journal\fi\hfill\today}{\relax}{}{}
\makeatother

\begin{document}

\begin{frontmatter}

\title{First interpretation of 13~TeV supersymmetry searches in the pMSSM}

\author{Alan Barr}
\author{Jesse Liu\corref{cor1}}
\ead{jesse.liu@physics.ox.ac.uk}

\address{Department of Physics, University of Oxford, Denys Wilkinson Building,
 Oxford OX1 3RH, United Kingdom
}

\cortext[cor1]{Corresponding author}

\begin{abstract}
The combined constraints from six early Run~2 ATLAS searches for supersymmetry are interpreted in the phenomenological minimal supersymmetric extension to the Standard Model (pMSSM). Each of the searches was based on proton--proton collision data recorded in 2015 at $\sqrt{s} =13$~TeV with 3.2~fb$^{-1}$ of integrated luminosity. Sensitivity to squarks of the first two generations and gluinos are evaluated using fast detector simulation. Results are presented in the 19-parameter R-parity conserving pMSSM with the lightest supersymmetric particle (LSP) being the neutralino. Considering 181.8k points that survived Run~1 constraints, 15.7\% are excluded at 95\% confidence level. Of those satisfying these Run~2 constraints, 0.5\% (1.0\%) have sub-TeV gluinos (sub-500~GeV squarks), the lightest of which has a mass of 757~GeV (293~GeV) with a 689~GeV (217~GeV) LSP.   \textit{\footnotesize \linebreak \\ This is not the work of, nor endorsed by, the ATLAS or CMS collaborations.}
\end{abstract}

\begin{keyword}
LHC \sep supersymmetry \sep pMSSM
\end{keyword}

\patchcmd{\MaketitleBox}{\footnotesize\itshape\elsaddress\par\vskip36pt}{\footnotesize\itshape\elsaddress\par\parbox[b][26pt]{\linewidth}{\vfill\hfill\textnormal{\vspace{0.3cm}\today}\hfill\null\vfill}}{}{}%

\end{frontmatter}



\section{Introduction}

Direct searches for supersymmetry (SUSY) remain central to the LHC physics programme. Introducing supersymmetric particles (sparticles) ameliorates fine-tuning in the Higgs sector, facilitates unification of gauge couplings and, when \mbox{R-parity} is conserved, provides dark matter candidates. Naturalness arguments favour coloured sparticles to have weak-scale masses and to be within reach at the LHC. For Run~2 in 2015, the ATLAS and CMS collaborations resumed searches for supersymmetry using 3.2~fb$^{-1}$ and 2.2~fb$^{-1}$ of integrated luminosity respectively. However the dearth of statistically significant signals persists~\cite{Aaboud:2016zdn, Aad:2016jxj, Aad:2016qqk, Aad:2016multib, Aad:2016tuk, Aaboud:2016tnv, Khachatryan:2016xvy, Khachatryan:2016kdk, Khachatryan:2016kod, Khachatryan:2016uwr, Aad:2014wea, Aad:2013yna, Aad:2015iea, Aad:2015pfx, Aad:2015eda, Khachatryan:2015wza, Khachatryan:2015vra}. Experimental searches usually present lower mass bounds in terms of simplified models comprising a small number of kinematically accessible sparticles~\cite{Alwall:2008ag}, or small subspaces of the minimal supersymmetric extension to the Standard Model (MSSM) derived from theoretical assumptions at high-energy scales~\cite{Kane:1993td, Giudice:1998bp}. 

An alternative framework advocated for interpretation is the p(henomenological)MSSM~\cite{Berger:2008cq, CahillRowley:2012cb, CahillRowley:2012kx,Cahill-Rowley:2014twa}. The ATLAS collaboration recently re-examined 22 Run~1 searches at 7 and 8~TeV using up to 20.3~fb$^{-1}$ of integrated luminosity within a 19-parameter pMSSM~\cite{Aad:2015baa}. By employing the complete ATLAS detector simulation, reconstruction and analysis software, the collaboration performed a comprehensive assessment on the status of the MSSM after Run~1. The parametric freedom allowed them to explore correlations between independent searches and phenomenological implications on non-collider observables such as dark matter. They found scenarios with percent level fine-tuning that remained viable after Run~1 and within Run~2 reach. The rich exclusion information is both valuable for improving search strategies and also amenable for novel machine learning studies~\cite{Caron:2016hib}, further meriting this broad interpretation approach.

This Letter presents the first interpretation of ATLAS searches based on data taken at $\sqrt{s}=13$~TeV in 2015 within the 19-parameter pMSSM. Run~2 sensitivity already extends simplified model limits beyond their 8 TeV counterparts and is equally expected to impact this pMSSM space. Using fast detector simulation, we interpret constraints from six Run~2 searches on the set of pMSSM points considered by ATLAS in Run 1~\cite{Berger:2008cq, CahillRowley:2012cb, CahillRowley:2012kx,Cahill-Rowley:2014twa, Aad:2015baa}. In this work, we assess sensitivity to strongly interacting sparticles, given available luminosity, specifically production of squarks of the first two generations and gluinos.
\section{Model points and methodology}


\subsection{Review of ATLAS pMSSM}

The ATLAS collaboration obtained their 19-parameter pMSSM points by making several phenomenologically motivated assumptions; see Ref.~\cite{Aad:2015baa} for full details. The discrete symmetry R-parity was taken to be exactly conserved, and the lightest supersymmetric particle (LSP) was required to be the neutralino~$\tilde{\chi}_1^0$. Minimal flavour violation was imposed and no new CP violating phases were introduced. The resulting 19 free parameters were scanned with flat priors, subject to LEP lower bounds on sparticle masses. The upper ceiling on mass scales was taken to be 4~TeV. Indirect constraints from precision electroweak measurements, muon $g_\mu-2$ , $Z^0$ invisible width, Higgs mass, heavy flavour physics were then applied to over 3 million resulting models. The LSP was not assumed to be the sole constituent of cosmic dark matter so only the Planck~\cite{Ade:2015xua} upper bound was placed on the neutralino relic density $\Omega_{\tilde{\chi}_1^0}h^2$.

The remaining 310.3k models were evaluated by ATLAS against 22 Run~1 searches, as described in Ref.~\cite{Aad:2015baa}. Models with sensitivity were excluded at 95\% confidence level. ATLAS categorised models by the dominant contribution of the LSP being the bino, wino or Higgsino as defined in table~\ref{tab:LSPtype}, due to their different resulting phenomenology. Overall, ATLAS found 40.9\% of models to be excluded by their Run 1 analyses, with the \mbox{2--6 jets} search~\cite{Aad:2014wea} being most constraining. Searches for coloured sparticles had greatest sensitivity and reasonable corroboration with simplified models was reported. ATLAS noted the disappearing track search~\cite{Aad:2013yna} was also highly constraining, especially to wino-like LSP models with metastable charginos. The distributions of sparticle masses were altered, most substantially at lower masses, as one would expect.

\begin{table}[]
\centering
    \begin{tabular}{lccc}
    \toprule 
     LSP type & Definition  \\ \midrule
    `Bino-like' $\tilde{B}$     & $N_{11}^2 > \textrm{max}\left(N_{12}^2, N_{13}^2 + N_{14}^2\right)$ \\
    `Wino-like' $\tilde{W}$     & $N_{12}^2 > \textrm{max}\left(N_{11}^2, N_{13}^2 + N_{14}^2\right)$  \\
    `Higgsino-like' $\tilde{H}$ & $\left(N_{13}^2 + N_{14}^2\right) > \textrm{max}\left(N_{11}^2, N_{12}^2 \right)$\\
    \bottomrule
    \end{tabular}
\caption{\label{tab:LSPtype} Definition of neutralino $\tilde{\chi}^0_1$ LSP categories from Ref.~\cite{Aad:2015baa}. In the neutralino mixing parameter $N_{ij}$, the first index denotes the neutralino mass eigenstate $\tilde{\chi}^0_i$ and the second indicates its nature in the order $\left(\tilde{B}, \tilde{W}, \tilde{H}_1, \tilde{H}_2\right)$.}
\end{table}

\begin{table}[]
\centering
    \begin{tabular}{lccc}
    \toprule 
     Models & Bino & Wino & Higgsino \\ \midrule
    Viable after ATLAS Run 1 & 61.6k & 43.8k & 78.4k\\
    Without long-lived & 59.9k & 43.6k & 78.3k \\
    Without LL, with $\sigma_{\rm tot} \geq 5$~fb                        & 
    48.7k      & 29.7k & 52.8k  \\ \bottomrule
    \end{tabular}
\caption{\label{tab:generatedfrac} Viable model points before Run~2 constraints. These are classified by the dominant contribution to the LSP being bino, wino or Higgsino. Long-lived (LL) models with $c\tau > 1$~mm require dedicated Monte-Carlo simulation and are omitted from this study. Event simulation was performed on non-LL models with total strong sparticle production cross-section $\sigma_{\rm tot} \geq 5$~fb.  }
\end{table}

\subsection{13~TeV signal and detector simulation}

The 183.8k models from Ref.~\cite{Aad:2015baa} that survived Run~1 constraints are considered for 13~TeV sensitivity with the following procedure. First, we omit 1\% of models featuring long-lived sparticles with $c\tau > 1$~mm (as defined in Ref.~\cite{Aad:2015baa}), since they require dedicated simulation beyond the scope of this Letter. Next, we calculate the 13~TeV total production cross-sections $\sigma_{\rm tot}$ of any two coloured sparticles at leading-order using \texttt{MadGraph5}~v2.3.3~\cite{Alwall:2011uj, Alwall:2014hca}. Based on sample studies, we assume the subset of models with $\sigma_{\rm tot}$ smaller than 5~fb have no sensitivity with 3.2~fb$^{-1}$ of Run 2 luminosity and are deemed not excluded.
The bottom row of table~\ref{tab:generatedfrac} consists of the remaining models selected to perform signal and detector simulation as follows. 

We use \texttt{MadGraph5} to generate events in which any two coloured sparticles are produced from proton-proton collisions, with up to one additional parton in the matrix element. These events are computed at tree level then showered and hadronised by \texttt{Pythia}~6.428~\cite{Sjostrand:2006za}, employing the \texttt{CTEQ6L} parton distribution functions~\cite{Pumplin:2002vw}. The MLM prescription~\cite{Mangano:2006rw} is used to match jets with the \texttt{MadGraph} minimum parton $k_T$ parameter set to 100~GeV and \texttt{Pythia} jet measure cutoff at 120~GeV, in accord with Ref.~\cite{Aad:2015baa}. 
The  \texttt{Delphes}~3.3.2~\cite{deFavereau:2013fsa} package performs fast detector simulation, using the anti-$k_T$ clustering algorithm with cone parameter $R = 0.4$ in the \texttt{Fastjet}~3.1.3 package~\cite{Cacciari:2008gp, Cacciari:2011ma} for jet reconstruction. The exact event cleaning requirements, object reconstruction efficiencies, and isolation and overlap criteria vary between analyses. To reduce disk usage to a manageable level, we take these quantities to be universal and produce one \texttt{Delphes} file of reconstructed objects per pMSSM point. This parametrisation yields acceptable results during validation and we manually reweight events at the analysis stage where necessary.

\begin{table}[]
\centering
\begin{tabular}{lcccc}
\toprule
  Analysis   & \multicolumn{1}{c}{All LSPs} & \multicolumn{1}{c}{Bino} & \multicolumn{1}{c}{Wino} & \multicolumn{1}{c}{Higgsino} \\ \midrule
2--6 jets~\cite{Aaboud:2016zdn}     &12.6\%     & 17.2\%    & 10.8\%    & 10.1\% \\ 
7--10 jets~\cite{Aad:2016jxj}       &0.6\%      & 0.5\%     & 0.5\%     & 0.7\%\\
1-lepton~\cite{Aad:2016qqk} &1.0\%      & 0.8\%     & 1.1\%     & 1.1\%\\
Multi-b~\cite{Aad:2016multib}  &4.2\%      & 3.0\%     & 4.0\%     & 5.2\%\\
SS/3L~\cite{Aad:2016tuk}            &0.5\%      & 0.1\%     & 1.6\%     & 0.1\%\\ 
Monojet~\cite{Aaboud:2016tnv}       &1.3\%      & 3.3\%     & 0.2\%     & 0.2\%\\ \midrule
 All analyses                       &15.7\%     & 18.8\%    & 14.9\%    & 13.8\% \\ \bottomrule
\end{tabular}
\caption{\label{tab:listSearches} Percentage of model points excluded by the considered Run~2 analyses out of the points that survived Run~1 constraints, with a breakdown by LSP type. All percentages are normalised to the number of models satisfying `Without long-lived' in table~\ref{tab:generatedfrac}.
}
\end{table}

\subsection{Recasting ATLAS analyses}

To replicate ATLAS event selection, we utilise the \texttt{MadAnalysis5}~v1.3 recasting framework~\cite{Conte:2012fm, Conte:2014zja}. We adapt codes from the MadAnalysis Public Analysis Database (PAD)~\cite{Dumont:2014tja, MArun1ATLAS:2-6jets, MArun1ATLAS:7+jets} for the 13~TeV ATLAS analyses listed in table~\ref{tab:listSearches} where available, and write our own otherwise. ATLAS optimised the searches to target production of squarks and gluinos, typically decaying into jets accompanied by missing transverse momentum with magnitude $E_T^{\rm miss}$. The \mbox{2--6 jets} and \mbox{1-lepton} analyses apply the aplanarity variable~\cite{Aad:2012np, Chen:2011ah}. We implement top-tagging in our \mbox{Multi-b} analysis by reclustering $R=0.4$ jets using the anti-$k_T$ algorithm into $R=1.0$ jets and retaining those that satisfy the kinematic criteria defined in Ref.~\cite{Aad:2016multib}. Full details of event selection are in the ATLAS references. 

Each of our implemented analyses are validated to ensure consistent behaviour with the corresponding ATLAS search. Cutflows for at least one benchmark point per analysis have better than 30\% agreement at each event selection, which is within systematic uncertainties. We also reproduce the observed simplified model limits to within 100~GeV of those published by ATLAS, utilising next-to-leading order (NLO) cross-sections from Ref.~\cite{Borschensky:2014cia}. After samples are generated and analysed, the \texttt{RecastingTools} package in \texttt{MadAnalysis5} performs statistical tests using the CL$_{\rm s}$ prescription~\cite{Read:2002hq}. We deem a pMSSM point excluded if the best expected CL$_s$ value is less than 0.05 for at least one analysis. No attempt is made to statistically combine results from different analyses, which is beyond the scope of this study. While we would typically expect NLO cross-sections to systematically increase signal yields by order 30\%, this is within the considerable parton distribution function uncertainties. Thus leading-order cross-sections we calculated in \texttt{MadGraph5} are used together with 3.2~fb$^{-1}$ of integrated luminosity to normalise event yields for each pMSSM point. 
\section{Results and discussion}


\subsection{Exclusion by Run~2 analyses}

Table~\ref{tab:listSearches} displays the fraction of model points excluded by each Run~2 analysis out of those points that survived Run~1 constraints, with a breakdown by LSP type. All such fractions are normalised to the number of models satisfying `Without long-lived' in table~\ref{tab:generatedfrac}. Overall, 15.7\% of models surviving Run~1 are excluded by Run~2 analyses considered by us. Bino-like LSP models are more constrained at 18.8\% compared with 14.9\% and 13.8\% for wino and Higgsino counterparts respectively. Consistent with Ref.~\cite{Aad:2015baa}, the \mbox{2--6 jets} search remained the most constraining at Run 2. Excluding 12.6\% of all models analysed, it was particularly sensitive to bino-like LSPs. We find the next most constraining analysis is the \mbox{Multi-b}, excluding 4.2\% of models.

This larger fraction of bino-like LSP models being excluded arises from the prior distribution of gluinos. Before applying our analyses, gluino masses were generally skewed towards higher masses due to stringent Run~1 constraints. However for bino-like LSP models, a greater proportion of models with sub-TeV gluinos remain. This can be understood in terms of cosmological relic density requirements. Models with bino-like LSPs typically oversaturate the Planck dark matter bound and thus require a near-degenerate next-to-LSP (NLSP) such as a gluino to act as an early universe co-annihilator. 

Table~\ref{tab:overlap} shows the percentage of models excluded by an analysis in each row that is also excluded by another in the columns. Though several searches were optimised for pair gluino production, they targeted different decay processes or final states such as lepton presence. The good complementarity between different searches is therefore exhibited for Run~2, with varying degrees of overlap. For example, of the models excluded by the \mbox{2--6 jets} search, 13\% are also excluded by the \mbox{Multi-b} analysis. The \mbox{2--6 jets} search also features a `2jm' signal region, which targets sparticles recoiling off an energetic jet from initial-state radiation, similar to the dedicated Monojet analysis. Though the resulting overlap is large, the Monojet search maintains unique sensitivity to $0.2\%$ of models featuring very small mass splittings between the produced sparticle and LSP. 

\begin{table}[]
\centering
\resizebox{0.48\textwidth}{!}{%
\begin{tabular}{lcccccc}
\toprule
           & 2--6 jets & 7--10 jets & 1-lepton & Multi-b & SS/3L  & Monojet \\
2--6 jets  & 100\%     & 3\%        & 5\%      & 13\%    & 0\%    & 10\% \\
7--10 jets & 76\%      & 100\%      & 59\%     & 91\%    & 4\%    & 6\% \\
1-lepton   & 65\%      & 34\%       & 100\%    & 55\%    & 8\%    & 7\% \\
Multi-b    & 39\%      & 12\%       & 13\%     & 100\%   & 1\%    & 1\% \\
SS/3L      & 10\%      & 5\%        & 17\%     & 6\%     & 100\%  & 3\% \\ 
Monojet    & 99\%      & 3\%        & 6\%      & 5\%     & 1\%    & 100\% \\ \bottomrule
\end{tabular}
}
\caption{\label{tab:overlap} Exclusion overlap: percentage of models excluded by a Run~2 analysis on each row that is also excluded by another in the columns. 100\% is reserved for complete overlap of models excluded. 
}
\end{table}

\subsection{Gluino and squark masses}

\begin{figure*}[]
    \centering
         \includegraphics[width=0.49\textwidth]{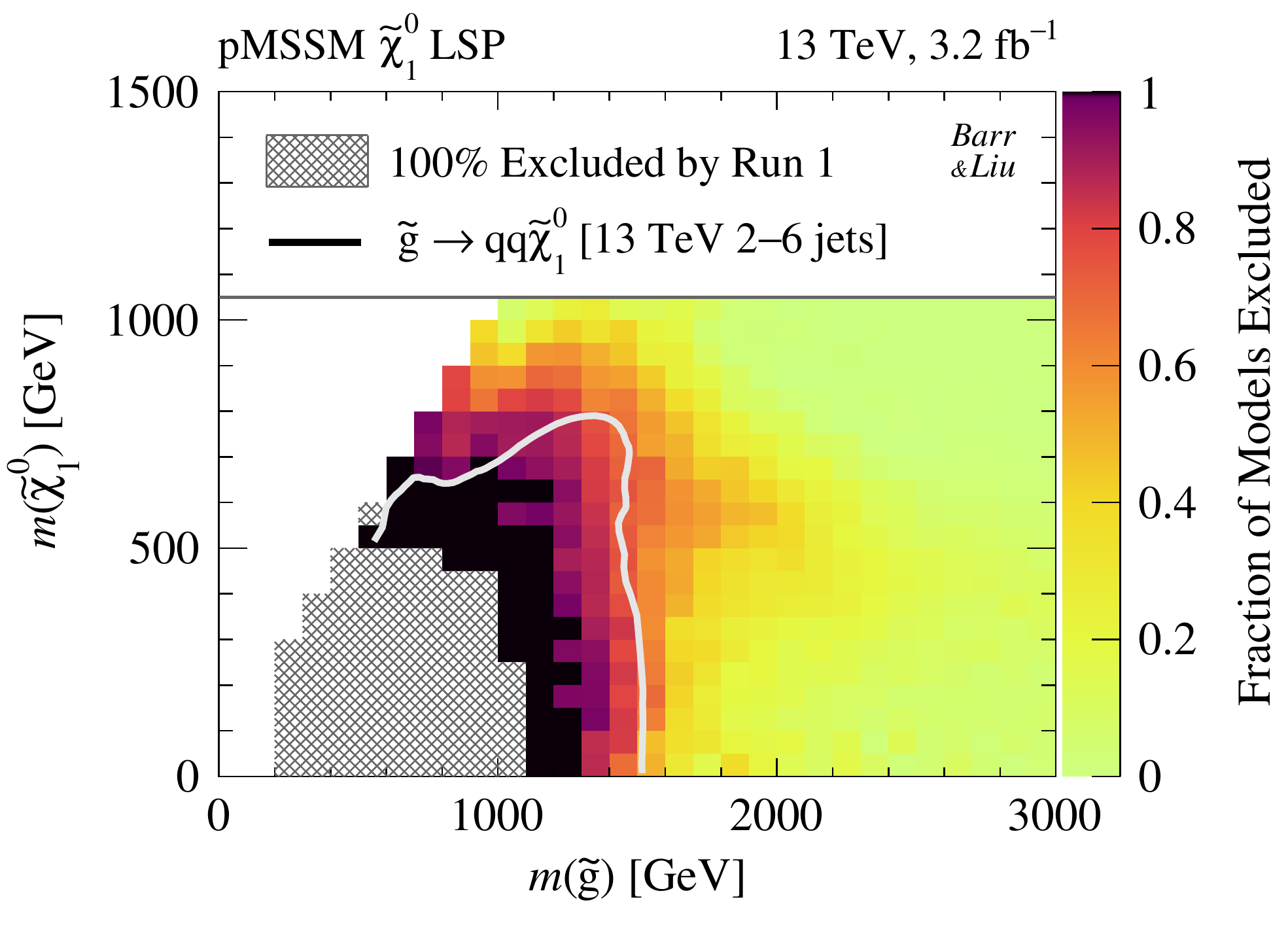}
         \includegraphics[width=0.49\textwidth]{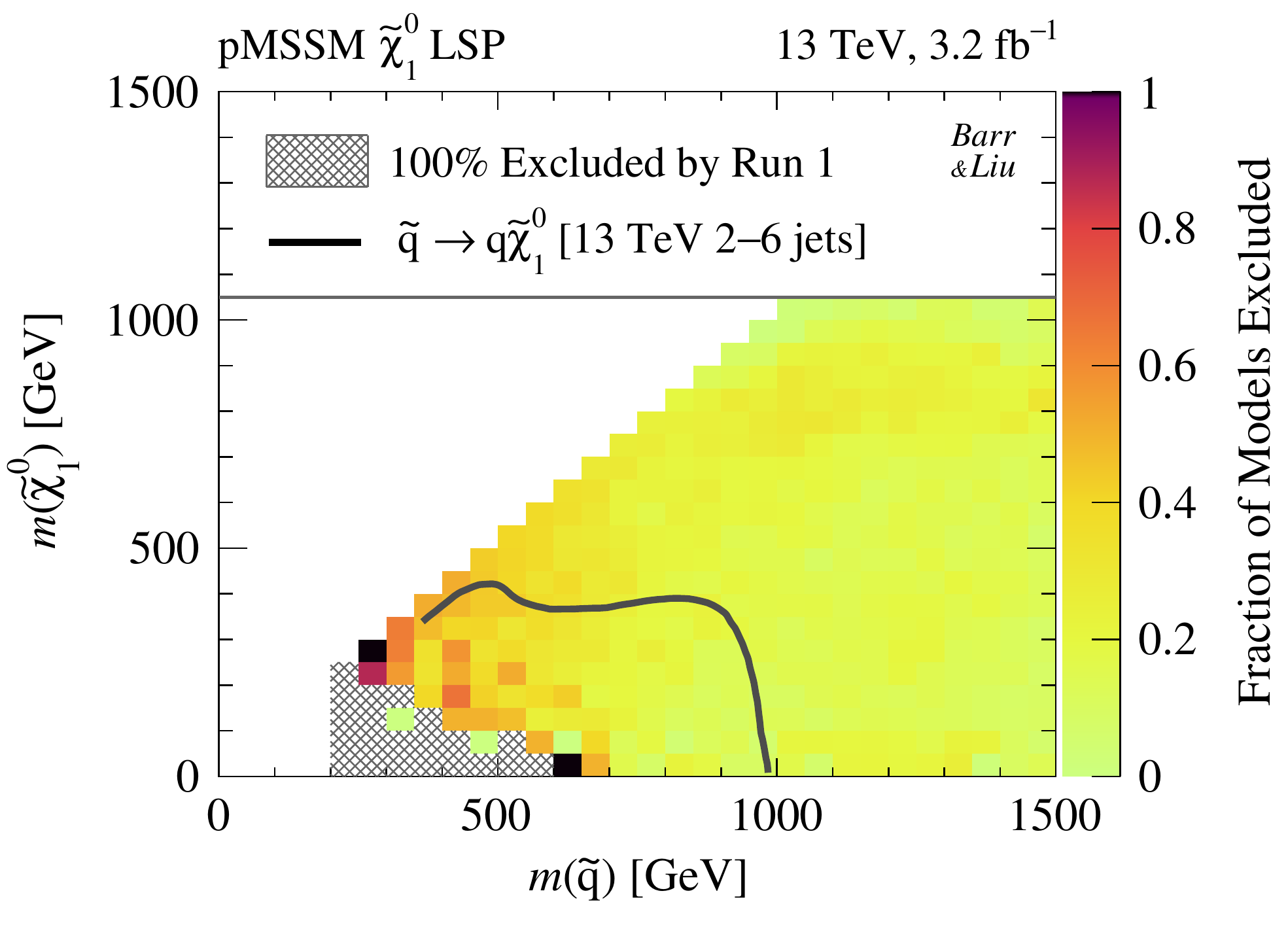}
    \caption{Fraction of model points excluded by the combined constraints from the six early Run~2 searches considered in table~\ref{tab:listSearches} out of the points that survived Run~1 constraints. In each mass bin, the colour scale denotes the fraction of points excluded at 95\% confidence level normalised to the number of points satisfying `Without long-lived' in table~\ref{tab:generatedfrac}, where black indicates 100\% exclusion.  This is projected into the gluino--LSP $\tilde{g}$-$\tilde{\chi}_1^0$ (left) and squark--LSP $\tilde{q}$-$\tilde{\chi}_1^0$ (right) mass planes. Here $m(\tilde{q})$ is the mass of the lightest squark among the first two generations. No models were produced by ATLAS in white regions due to non-LHC constraints. Hatched grey regions indicate that all models were excluded by Run~1 in Ref.~\cite{Aad:2015baa}. Overlayed grey solid lines are the `direct decay' simplified model limits from the 13~TeV 2--6 jets search~\cite{Aaboud:2016zdn} for gluinos $\tilde{g} \to qq\tilde{\chi}^0_1$ (left) and squarks $\tilde{q} \to q\tilde{\chi}^0_1$ (right). In the latter case, all eight squarks are of the first two generations are assumed to be mass-degenerate. 
    }
    \label{fig:mass_plane_Gl_Sqk_chi10}
\end{figure*}

Results are also presented by projecting the 19-dimensional pMSSM into two-parameter subspaces. While comparisons can be made with the ATLAS summary paper~\cite{Aad:2015baa}, care must be taken with interpretation. The parameter space is highly correlated, with some regions being particularly sensitive to indirect constraints. Moreover, we display fractions of models excluded normalised to those that survived Run 1 constraints, with long-lived models removed. This fraction can be sensitive to large changes in prior distributions of sparticle mass compared with Ref.~\cite{Aad:2015baa}. We focus on gluinos and squarks, where the latter henceforth refers to any of the mass states resulting from left- or right- handed dominated squarks of either the first or second generation $\tilde{q}\in \{\tilde{d}, \tilde{u}, \tilde{s}, \tilde{c}\}_{L,R}$.  

Figure~\ref{fig:mass_plane_Gl_Sqk_chi10} displays the fraction of model points excluded by our combination of the Run~2 analyses out of the points that survived Run~1 constraints, projected into the gluino--LSP $\tilde{g}$-$\tilde{\chi}_1^0$ and lightest squark--LSP $\tilde{q}$-$\tilde{\chi}_1^0$ mass planes. The increased sensitivity of Run~2 searches is already unambiguous, especially at low sparticle masses. Production cross-sections and therefore sensitivity decrease with higher masses. Gluinos experience a greater boost in sensitivity going from 8 to 13~TeV than squarks due to advantageous scaling of LHC parton distribution functions. There is good corroboration with the simplified model limit of gluinos directly decaying to LSP $\tilde{g} \to qq\tilde{\chi}_1^0$ from the \mbox{2--6 jets} search. With richer mass spectra in the pMSSM, cascade decays can often be favoured when other sparticles exist between the gluino and LSP, modifying jet kinematics, $E_T^{\rm miss}$ signatures and consequently sensitivity.   

\begin{figure}[h!]
    \centering
         \includegraphics[width=0.48\textwidth]{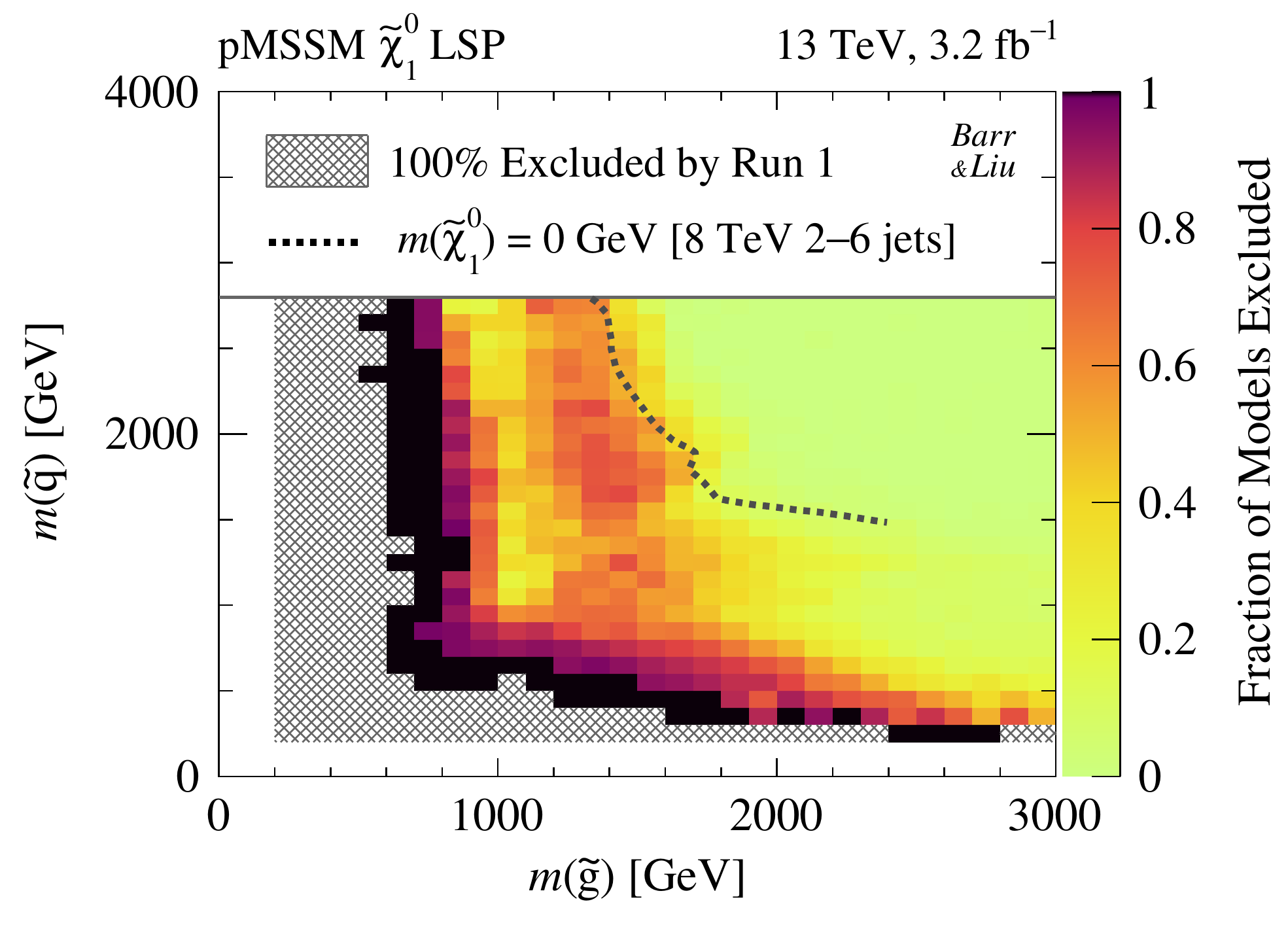}
    \caption{Fraction of model points excluded by the Run~2 searches considered in table~\ref{tab:listSearches} out of the points that survived Run~1 constraints, projected into the gluino--squark $\tilde{g}$-$\tilde{q}$ mass plane. The overlayed grey dashed line is the `gluino--squark--LSP simplified pMSSM' scenario from the 8 TeV 2--6 jets search \cite{Aad:2014wea}. Here $m(\tilde{q})$ is the mass of the lightest squark among the first two generations. The colour scheme is described in figure~\ref{fig:mass_plane_Gl_Sqk_chi10}.}
    \label{fig:mass_plane_Gl_Sqk}
\end{figure}

Similar observations apply to the squark--LSP plane in figure~\ref{fig:mass_plane_Gl_Sqk_chi10}~(right). Moving to 13~TeV, sensitivity to squarks increases but less substantially than for gluinos, as evident in the figure. Nevertheless exclusion, especially at low squark masses, already extends beyond Run~1. The direct squark $\tilde{q} \to q\tilde{\chi}^0_1$ simplified model assumes the lightest squark is mass degenerate with all other left- or right- handed first or second generation squarks. The exclusion of high gluino/squark masses is understood to be a result of mass correlations between different sparticles allowed by the pMSSM. Figure~\ref{fig:mass_plane_Gl_Sqk} illustrates this by projecting into the mass plane of lightest squark vs gluino. Models with high gluino masses can be excluded due to the presence of low mass squarks, and vice versa. This projection also reveals a localised region of high exclusion around 1.4~TeV gluino and 1.8~TeV squark masses. We find this is largely due to the \mbox{Multi-b} search rather than the \mbox{2--6} jets analysis, again demonstrating the complementarity of analyses. For this projection, the most recent ATLAS simplfied model available for comparison is from the Run 1 8~TeV 2--6 jets search~\cite{Aad:2014wea}. 

Of the model points that survive our Run~2 constraints, 0.5\% contain sub-TeV gluinos, the lightest of which is model number 189200115 with a 757~GeV gluino and 689~GeV LSP. Meanwhile 1.0\% have at least one sub-500~GeV squark, with model number 8243590 featuring the lightest squark mass of 293~GeV accompanied by a 217~GeV LSP. Furthermore, ATLAS identified models with lowest fine-tuning (percent level) favoured low third generation squark masses, with much of the other spectrum decoupled~\cite{Aad:2015baa}. These remain viable after the early Run~2 papers analysed here, though it should be noted that third generation squarks were beyond the scope of this study.

\section{Conclusion}

In summary, this study interpreted results from six early Run~2 ATLAS searches for supersymmetry to assess its impact on the 19 parameter R-parity conserving pMSSM with neutralino LSP. Each analysis was based on LHC \mbox{proton--proton} collision data at 13~TeV with 3.2~fb$^{-1}$ of integrated luminosity recorded in 2015. Considering 181.8k model points that survived Run~1 constraints, 71.4\% underwent particle and fast detector simulation, with over 3 billion events generated. We considered sensitivity to squarks of the first two generations and gluinos. 

Of the pMSSM points that survived Run~1 constraints, $15.7\%$ were excluded at 95\% confidence level by Run~2 analyses. This demonstrates the considerable sensitivity of Run~2 searches beyond Run~1, being particularly salient for low-mass gluinos. The 2--6 jets analysis was the most constraining, excluding 12.6\% models alone, with the next most constraining being the Multi-b search at 4.2\%. The good complementarity between searches was also exhibited. Of the model points satisfying Run~2 constraints, 0.5\% (1.0\%) contain sub-TeV gluinos (sub-500~GeV squarks), the lightest of which has a mass of 757~GeV (293~GeV) accompanied by a 689~GeV (217~GeV) LSP. With greater luminosity, LHC Run~2 is expected to bring such points to sensitivity and will continue to probe hitherto unexplored regions of the pMSSM space.

\section*{Acknowledgements}
We are grateful to Will Fawcett for invaluable conversations, especially on the use of \texttt{SLHA} model cards and clarifying the consideration of long-lived models. Further thanks go to Claire Gwenlan, Will Kalderon and Mike Nelson for helpful discussions, together with the computing support of Dennis Liu and Kashif Mohammad. This research is supported by STFC. The authors would like to acknowledge the use of the University of Oxford Advanced Research Computing (ARC) facility in carrying out this work \cite{richards_2015_22558}.





\bibliographystyle{atlas-style-no-title}


\bibliography{./bib/all-refs}

\providecommand{\href}[2]{#2}\begingroup\raggedright\begin{thebibliography}{10}

\bibitem{Aaboud:2016zdn}
{ATLAS} Collaboration,
\href{http://arxiv.org/abs/1605.03814}{{\ttfamily arXiv:1605.03814}}.

\bibitem{Aad:2016jxj}
{ATLAS} Collaboration,
  \href{http://dx.doi.org/10.1016/j.physletb.2016.04.005}{Phys. Lett.
  {\bfseries B757} (2016) 334--355},
\href{http://arxiv.org/abs/1602.06194}{{\ttfamily arXiv:1602.06194}}.

\bibitem{Aad:2016qqk}
{ATLAS} Collaboration,
\href{http://arxiv.org/abs/1605.04285}{{\ttfamily arXiv:1605.04285}}.

\bibitem{Aad:2016multib}
{ATLAS} Collaboration,
\href{http://arxiv.org/abs/1605.09318}{{\ttfamily arXiv:1605.09318}}.

\bibitem{Aad:2016tuk}
{ATLAS} Collaboration,
  \href{http://dx.doi.org/10.1140/epjc/s10052-016-4095-8}{Eur. Phys. J.
  {\bfseries C76} no.~5, (2016) 259},
\href{http://arxiv.org/abs/1602.09058}{{\ttfamily arXiv:1602.09058}}.

\bibitem{Aaboud:2016tnv}
{ATLAS} Collaboration,
\href{http://arxiv.org/abs/1604.07773}{{\ttfamily arXiv:1604.07773}}.

\bibitem{Khachatryan:2016xvy}
{CMS} Collaboration,
\href{http://arxiv.org/abs/1603.04053}{{\ttfamily arXiv:1603.04053}}.

\bibitem{Khachatryan:2016kdk}
{CMS} Collaboration,
  \href{http://dx.doi.org/10.1016/j.physletb.2016.05.002}{Phys. Lett.
  {\bfseries B758} (2016) 152--180},
\href{http://arxiv.org/abs/1602.06581}{{\ttfamily arXiv:1602.06581}}.

\bibitem{Khachatryan:2016kod}
{CMS} Collaboration,
\href{http://arxiv.org/abs/1605.03171}{{\ttfamily arXiv:1605.03171}}.

\bibitem{Khachatryan:2016uwr}
{CMS} Collaboration,
\href{http://arxiv.org/abs/1605.04608}{{\ttfamily arXiv:1605.04608}}.

\bibitem{Aad:2014wea}
{ATLAS} Collaboration, \href{http://dx.doi.org/10.1007/JHEP09(2014)176}{JHEP
  {\bfseries 09} (2014) 176},
\href{http://arxiv.org/abs/1405.7875}{{\ttfamily arXiv:1405.7875}}.

\bibitem{Aad:2013yna}
{ATLAS} Collaboration,
  \href{http://dx.doi.org/10.1103/PhysRevD.88.112006}{Phys. Rev. {\bfseries
  D88} no.~11, (2013) 112006},
\href{http://arxiv.org/abs/1310.3675}{{\ttfamily arXiv:1310.3675}}.

\bibitem{Aad:2015iea}
{ATLAS} Collaboration, \href{http://dx.doi.org/10.1007/JHEP10(2015)054}{JHEP
  {\bfseries 10} (2015) 054},
\href{http://arxiv.org/abs/1507.05525}{{\ttfamily arXiv:1507.05525}}.

\bibitem{Aad:2015pfx}
{ATLAS} Collaboration,
  \href{http://dx.doi.org/10.1140/epjc/s10052-015-3726-9}{Eur. Phys. J.
  {\bfseries C75} no.~10, (2015) 510},
\href{http://arxiv.org/abs/1506.08616}{{\ttfamily arXiv:1506.08616}}.

\bibitem{Aad:2015eda}
{ATLAS} Collaboration,
\href{http://arxiv.org/abs/1509.07152}{{\ttfamily arXiv:1509.07152}}.

\bibitem{Khachatryan:2015wza}
{CMS} Collaboration, \href{http://dx.doi.org/10.1007/JHEP06(2015)116}{JHEP
  {\bfseries 06} (2015) 116},
\href{http://arxiv.org/abs/1503.08037}{{\ttfamily arXiv:1503.08037}}.

\bibitem{Khachatryan:2015vra}
{CMS} Collaboration, \href{http://dx.doi.org/10.1007/JHEP05(2015)078}{JHEP
  {\bfseries 05} (2015) 078},
\href{http://arxiv.org/abs/1502.04358}{{\ttfamily arXiv:1502.04358}}.

\bibitem{Alwall:2008ag}
J.~Alwall, P.~Schuster,  and N.~Toro,
  \href{http://dx.doi.org/10.1103/PhysRevD.79.075020}{Phys. Rev. {\bfseries
  D79} (2009) 075020},
\href{http://arxiv.org/abs/0810.3921}{{\ttfamily arXiv:0810.3921}}.

\bibitem{Kane:1993td}
G.~L. Kane, C.~F. Kolda, L.~Roszkowski,  and J.~D. Wells,
  \href{http://dx.doi.org/10.1103/PhysRevD.49.6173}{Phys. Rev. {\bfseries D49}
  (1994) 6173--6210},
\href{http://arxiv.org/abs/hep-ph/9312272}{{\ttfamily arXiv:hep-ph/9312272}}.

\bibitem{Giudice:1998bp}
G.~F. Giudice and R.~Rattazzi,
  \href{http://dx.doi.org/10.1016/S0370-1573(99)00042-3}{Phys. Rept. {\bfseries
  322} (1999) 419--499},
\href{http://arxiv.org/abs/hep-ph/9801271}{{\ttfamily arXiv:hep-ph/9801271}}.

\bibitem{Berger:2008cq}
C.~F. Berger, J.~S. Gainer, J.~L. Hewett,  and T.~G. Rizzo,
  \href{http://dx.doi.org/10.1088/1126-6708/2009/02/023}{JHEP {\bfseries 02}
  (2009) 023},
\href{http://arxiv.org/abs/0812.0980}{{\ttfamily arXiv:0812.0980}}.

\bibitem{CahillRowley:2012cb}
M.~W. Cahill-Rowley, J.~L. Hewett, S.~Hoeche, A.~Ismail,  and T.~G. Rizzo,
  \href{http://dx.doi.org/10.1140/epjc/s10052-012-2156-1}{Eur. Phys. J.
  {\bfseries C72} (2012) 2156},
\href{http://arxiv.org/abs/1206.4321}{{\ttfamily arXiv:1206.4321}}.

\bibitem{CahillRowley:2012kx}
M.~W. Cahill-Rowley, J.~L. Hewett, A.~Ismail,  and T.~G. Rizzo,
  \href{http://dx.doi.org/10.1103/PhysRevD.88.035002}{Phys. Rev. {\bfseries
  D88} no.~3, (2013) 035002},
\href{http://arxiv.org/abs/1211.1981}{{\ttfamily arXiv:1211.1981}}.

\bibitem{Cahill-Rowley:2014twa}
M.~Cahill-Rowley, J.~L. Hewett, A.~Ismail,  and T.~G. Rizzo,
  \href{http://dx.doi.org/10.1103/PhysRevD.91.055002}{Phys. Rev. {\bfseries
  D91} no.~5, (2015) 055002},
\href{http://arxiv.org/abs/1407.4130}{{\ttfamily arXiv:1407.4130}}.

\bibitem{Aad:2015baa}
{ATLAS} Collaboration, \href{http://dx.doi.org/10.1007/JHEP10(2015)134}{JHEP
  {\bfseries 10} (2015) 134},
\href{http://arxiv.org/abs/1508.06608}{{\ttfamily arXiv:1508.06608}}.

\bibitem{Caron:2016hib}
S.~Caron, J.~S. Kim, K.~Rolbiecki, R.~R. de~Austri,  and B.~Stienen,
\href{http://arxiv.org/abs/1605.02797}{{\ttfamily arXiv:1605.02797}}.

\bibitem{Ade:2015xua}
{Planck} Collaboration, P.~A.~R. Ade {et~al.},
\href{http://arxiv.org/abs/1502.01589}{{\ttfamily arXiv:1502.01589}}.

\bibitem{Alwall:2011uj}
J.~Alwall, M.~Herquet, F.~Maltoni, O.~Mattelaer,  and T.~Stelzer,
  \href{http://dx.doi.org/10.1007/JHEP06(2011)128}{JHEP {\bfseries 06} (2011)
  128},
\href{http://arxiv.org/abs/1106.0522}{{\ttfamily arXiv:1106.0522}}.

\bibitem{Alwall:2014hca}
J.~Alwall, R.~Frederix, S.~Frixione, V.~Hirschi, F.~Maltoni, O.~Mattelaer,
  H.~S. Shao, T.~Stelzer, P.~Torrielli,  and M.~Zaro,
  \href{http://dx.doi.org/10.1007/JHEP07(2014)079}{JHEP {\bfseries 07} (2014)
  079},
\href{http://arxiv.org/abs/1405.0301}{{\ttfamily arXiv:1405.0301}}.

\bibitem{Sjostrand:2006za}
T.~Sjostrand, S.~Mrenna,  and P.~Z. Skands,
  \href{http://dx.doi.org/10.1088/1126-6708/2006/05/026}{JHEP {\bfseries 05}
  (2006) 026},
\href{http://arxiv.org/abs/hep-ph/0603175}{{\ttfamily arXiv:hep-ph/0603175}}.

\bibitem{Pumplin:2002vw}
J.~Pumplin, D.~R. Stump, J.~Huston, H.~L. Lai, P.~M. Nadolsky,  and W.~K. Tung,
  \href{http://dx.doi.org/10.1088/1126-6708/2002/07/012}{JHEP {\bfseries 07}
  (2002) 012},
\href{http://arxiv.org/abs/hep-ph/0201195}{{\ttfamily arXiv:hep-ph/0201195}}.

\bibitem{Mangano:2006rw}
M.~L. Mangano, M.~Moretti, F.~Piccinini,  and M.~Treccani,
  \href{http://dx.doi.org/10.1088/1126-6708/2007/01/013}{JHEP {\bfseries 01}
  (2007) 013},
\href{http://arxiv.org/abs/hep-ph/0611129}{{\ttfamily arXiv:hep-ph/0611129}}.

\bibitem{deFavereau:2013fsa}
{DELPHES 3} Collaboration, J.~de~Favereau, C.~Delaere, P.~Demin, A.~Giammanco,
  V.~Lemaître, A.~Mertens,  and M.~Selvaggi,
  \href{http://dx.doi.org/10.1007/JHEP02(2014)057}{JHEP {\bfseries 02} (2014)
  057},
\href{http://arxiv.org/abs/1307.6346}{{\ttfamily arXiv:1307.6346}}.

\bibitem{Cacciari:2008gp}
M.~Cacciari, G.~P. Salam,  and G.~Soyez,
  \href{http://dx.doi.org/10.1088/1126-6708/2008/04/063}{JHEP {\bfseries 04}
  (2008) 063},
\href{http://arxiv.org/abs/0802.1189}{{\ttfamily arXiv:0802.1189}}.

\bibitem{Cacciari:2011ma}
M.~Cacciari, G.~P. Salam,  and G.~Soyez,
  \href{http://dx.doi.org/10.1140/epjc/s10052-012-1896-2}{Eur. Phys. J.
  {\bfseries C72} (2012) 1896},
\href{http://arxiv.org/abs/1111.6097}{{\ttfamily arXiv:1111.6097}}.

\bibitem{Conte:2012fm}
E.~Conte, B.~Fuks,  and G.~Serret,
  \href{http://dx.doi.org/10.1016/j.cpc.2012.09.009}{Comput. Phys. Commun.
  {\bfseries 184} (2013) 222--256},
\href{http://arxiv.org/abs/1206.1599}{{\ttfamily arXiv:1206.1599}}.

\bibitem{Conte:2014zja}
E.~Conte, B.~Dumont, B.~Fuks,  and C.~Wymant,
  \href{http://dx.doi.org/10.1140/epjc/s10052-014-3103-0}{Eur. Phys. J.
  {\bfseries C74} no.~10, (2014) 3103},
\href{http://arxiv.org/abs/1405.3982}{{\ttfamily arXiv:1405.3982}}.

\bibitem{Dumont:2014tja}
B.~Dumont, B.~Fuks, S.~Kraml, S.~Bein, G.~Chalons, E.~Conte, S.~Kulkarni,
  D.~Sengupta,  and C.~Wymant,
  \href{http://dx.doi.org/10.1140/epjc/s10052-014-3242-3}{Eur. Phys. J.
  {\bfseries C75} no.~2, (2015) 56},
\href{http://arxiv.org/abs/1407.3278}{{\ttfamily arXiv:1407.3278}}.

\bibitem{MArun1ATLAS:2-6jets}
G.~Chalons and D.~Sengupta, tech. rep., 2015.
\newblock \url{http://doi.org/10.7484/INSPIREHEP.DATA.UYT6.GFD9}.

\bibitem{MArun1ATLAS:7+jets}
B.~Fuks, M.~Blanke,  and I.~Galon, tech. rep., 2015.
\newblock \url{http://doi.org/10.7484/INSPIREHEP.DATA.STLS.SAMT}.

\bibitem{Aad:2012np}
{ATLAS} Collaboration,
  \href{http://dx.doi.org/10.1140/epjc/s10052-012-2211-y}{Eur. Phys. J.
  {\bfseries C72} (2012) 2211},
\href{http://arxiv.org/abs/1206.2135}{{\ttfamily arXiv:1206.2135}}.

\bibitem{Chen:2011ah}
C.~Chen, \href{http://dx.doi.org/10.1103/PhysRevD.85.034007}{Phys. Rev.
  {\bfseries D85} (2012) 034007},
\href{http://arxiv.org/abs/1112.2567}{{\ttfamily arXiv:1112.2567}}.

\bibitem{Borschensky:2014cia}
C.~Borschensky, M.~Kr{\"a}mer, A.~Kulesza, M.~Mangano, S.~Padhi, T.~Plehn,  and
  X.~Portell, \href{http://dx.doi.org/10.1140/epjc/s10052-014-3174-y}{Eur.
  Phys. J. {\bfseries C74} no.~12, (2014) 3174},
  \href{http://arxiv.org/abs/1407.5066}{{\ttfamily arXiv:1407.5066}}.
\url{https://twiki.cern.ch/twiki/bin/view/LHCPhysics/SUSYCrossSections}.

\bibitem{Read:2002hq}
A.~L. Read,
\href{http://dx.doi.org/10.1088/0954-3899/28/10/313}{J. Phys. {\bfseries G28}
  (2002) 2693--2704}.

\bibitem{richards_2015_22558}
A.~Richards, tech. rep., University of Oxford, 2015.
\newblock \url{http://dx.doi.org/10.5281/zenodo.22558}.

\end{thebibliography}\endgroup

\end{document}